\documentstyle[aps,prl,twocolumn,floats,epsfig]{revtex}
\begin{document}
\draft

\wideabs{
\title{Experimental evidence of a fractal dissipative regime in high-T$_{c}$
superconductors}

\author{Mladen Prester}
\address{Institute of Physics, P.O.B. 304,
HR-10 000 Zagreb, Croatia}
\date{\today}
\maketitle

\begin{abstract}
We report on our experimental evidence of a substantial geometrical ingredient
characterizing the problem of incipient dissipation in high-T$_{c}$ superconductors
(HTS):  high-resolution studies of differential resistance-current characteristics in
absence of magnetic field enabled us to identify and quantify the fractal dissipative
regime inside which the actual current-carrying medium is an object of fractal
geometry.  The discovery of a fractal regime proves the reality and consistency of
critical-phenomena scenario as a model for dissipation in inhomogeneous and disordered
HTS, gives the experimentally-based value of the relevant finite-size scaling exponent
and offers some interesting new guidelines to the problem of pairing mechanisms in
HTS.
\end{abstract}

}
\narrowtext
%

Local inhomogeneities characterize HTS both on nanoscopic \cite{mih,mes} (e.g., periodic
or aperiodic variation of local oxygen stoichiometry) and mesoscopic \cite{sust} (e.g.,
oxygen depleted grain boundaries) spatial scales.  While all the consequences on normal
and superconducting charge transport in the former case has not been entirely clarified
yet the case of grain boundaries is better understood:  at least in a broad range of
experimental parameters the supercurrent transport in polycrystalline samples relies on
`weak link network' (WLN), i.e., on mesoscopic superconducting islands interconnected by
Josephson interaction.  Although the transport features on nanoscopic scale may
significantly differ from those characterizing a rather simple WLN problem (e.g., local
inhomogeneities seem to give rise, as reviewed by Refs.  \onlinecite{mih,mes}, to
conducting stripes, clusters, wires or filaments which are, at least to some extent,
mobile, compared to predominantly static weak links) the experimental evidence in favor
of {\em a-b} plane Josephson junctions \cite{jun} indicates that, besides qualitative
similarities, the intrinsic and WLN transport are more closely related one to another
than it had been foreseen earlier.

Irrespective of the extent the processes at nano- and mesoscopic scales are related, the
problem of charge transport in WLN represents an autonomous subject of much interest due
to its relevance for general understanding of transport in heterogeneous systems and in
Josephson Junction Arrays (JJA) in particular \cite{jja}.  Focusing to the problem of
dissipation there are convincing arguments, particularly in absence (or in small)
magnetic fields, that the onset of dissipation is dominated rather by a phenomenon of
percolation than the dynamical features of flux lattice \cite{pas,pre1}.  In a
disordered-bonds (DB) model \cite{pre1,sust} the critical current $I_{c}$ characterizing
the dissipation onset reflects the connectivity threshold $p_{c}$ of classical
percolation networks \cite{sta,bun} (such that $p_{c}=p(I_{c})$) so that the
experimentally documented power-law-like current-voltage (I-V) characteristics can be
naturally interpreted as a current-induced but in essence traditional critical
phenomenon.  Consequently, I-V characteristics should also
reveal various manifestations of crossover between the relevant length scales known to
underly the critical behavior.  We show in this report that the latter crossover may be
detected and quantitatively investigated in experimental I-V curves.  In particular, we
claim that the I-V characteristics are generically composed of the three distinct
regimes:  a regime revealing no practical dissipation, a regime obeying conventional
correlation length scaling (homogeneous regime), and an intermediate regime obeying
finite-size scaling (fractal regime).  While the dissipation in the former regimes
has been already a subject to experimental reports and appropriate modeling
\cite{pre1,gra}, the experimental results concerning the fractal regime are reported
here for the first time.

The basis of the model is the idea that the increasing current applied to disordered WLN
decreases the fraction of Josephson-current-carrying bonds in a random manner.  Hence, the
applied current plays the role of random generator which in classical random electrical
networks changes the relative fractions of their components.  In DB model \cite{pre1,sust}
the elements and the relevance of this analogy has been studied in details.  Here we focus
to the problem of relevant length scales.  In analogous classical networks there are two
of them \cite{sta,bun}:  the correlation length $\xi$ (the representative size of growing
ramified clusters ) for $p$ away from $p_{c}$ and the sample size $L$ for $p$ close to
$p_{c}$ (``at criticality'').  
\begin{figure}[h!!!]
\begin{center}
\leavevmode
\epsfig{file = 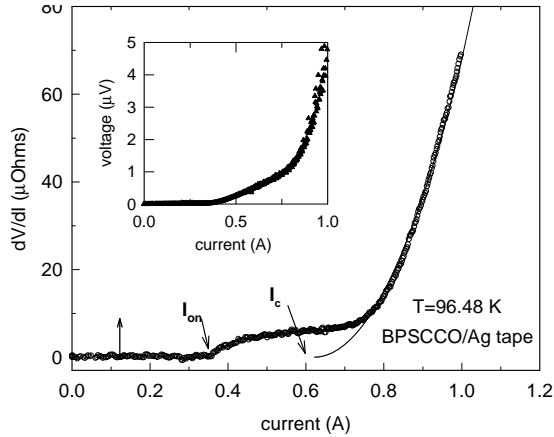, width =0.9 \linewidth, clip=}
\caption[delfsa]{High-resolution differential resistance in Ag/HTS composite tape, core
thickness $\approx 30 \mu m$.  The pronounced anomaly above the sharp dissipation onset
at $I_{on}$ is attributed to the sample size scaling ($\xi \geq L$).  For higher
currents the usual correlation length scaling ($\propto \xi^{-x/ \nu}$) takes over.
Experimentally, $dV/dI \propto (I-I_{c})^{x}, I_{c}=0.6, x=2$, as shown by the thin
line.  Vertical arrow illustrates the resolution:  Its length corresponds to $dc$
voltage of $15nV$.  The inset:  $V-I$ curve obtained by numerical integration of the
measured $dV/dI-I$.}
\label{f1}
\end{center}
\end{figure}
In approaching $p_{c}$, $\xi$ diverges involving exponent
$\nu$ ($ \approx 0.88$ in 3D systems) and the power-law form of static and dynamic
quantities manifests the property of spatial scaling.  In particular, the resistance $R$
of the random-superconductor network (RSN) disappears \cite{bun} as $(p_{c}-p)^{s} \propto
\xi ^{-s/ \nu}$, where $s$ is the breakdown exponent ($s \approx 0.8$ in 3D).  Close to
$p_{c}$ the homogeneous-to-fractal transformation of the geometry of incipient cluster
takes place and, while $R$ becomes independent of $p$, the finite-size scaling relation
\cite{bun,sta} $R(L) \propto L^{-s/ \nu-1}$ replaces the ordinary $R(L) \propto L^{-1}$
one. In applying a similar scenario to current-induced transition in WLN we assume that, in
approaching $I_{c}$ (i.e., $p_{c}$) from above, the representative size of the largest phase
coherent cluster diverges as $ \xi$ as well. The experimental studies \cite{pre1} of the
related homogeneous regime were shown to be in a close agreement with the predictions of the
model.  However, the precise interpretation \cite{fus} of the characteristic exponent
(experimentally, $dV/dI$ exponent is close to 2) is still unresolved (see discussion section).
Considering the experimental accessibility of the crossover to the fractal regime we note that
the unit of length involved in the WLN problem (i.e., average grain size $\ell$) belongs to
mesoscopic ($\mu m$) scale.  Hence, the observation of a size effect introduced by competing
length scales seems, for polycrystalline samples of reduced but still macroscopic thickness,
as an open possibility. Indeed, the first high resolution dynamical resistance measurements
\cite{pre2} (achieving the equivalent voltage resolution of better than 1nV) on
polycrystalline samples which are thin by their very design, HTS superconductor/normal metal
composite tapes (superconducting core thickness in the range 10-50 $\mu m$), revealed the two
characteristic currents.  As illustrated by Fig.\ \ref{f1}, the lower one triggers the onset
of low-level, non-exponential dissipation (onset current, $I_{on}$) while the higher one
parameterizes the scaling behavior, $dV/dI \propto (I-I_{c})^{x}, x\approx 2$, of subsequently
rapidly growing dissipation (thermodynamical critical current, $I_{c}$).  One could assume
that a rather broad dissipative range between $I_{on}$ and $I_{c}$ corresponds to validity of
$ \xi \geq L$ when the incipient dissipative sites would fill the sample-sized network of
fractal geometry. Indeed, the saturation-like behavior of dissipation in that range is, while
in obvious disagreement with any flux-creep model (exponential in applied current), at least
in qualitative agreement with general independence of any observable ($dV/dI$ in our case) on
$ \xi$ in the range of sample-sized fractal \cite{sta,bun}.  A similar observation of the
broad range of low-level dissipation in composite tapes has also been reported by other
authors but interpreted by less fundamental causes \cite{pol}.

\begin{figure}[h!!!]  
\begin{center} 
\leavevmode
\epsfig{file = 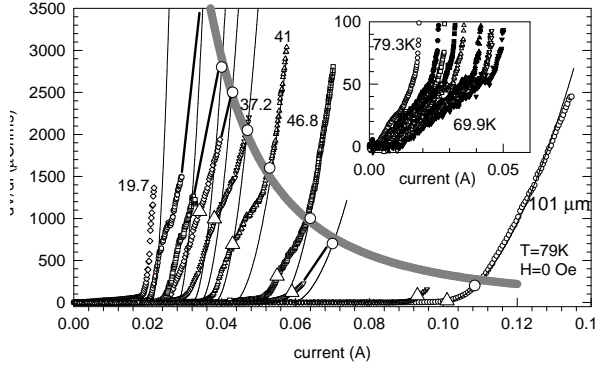, width =0.9 \linewidth, clip=} 
\caption[delfsa]{Thickness dependence
of differential resistances of polycrystalline GdBa$_{2}$Cu$_{3}$O$_{7-x}$ sample (open symbols)
and homogeneous-to-fractal phase boundary (thick grey line).  The thickness (in $\mu m$) is
designated by numbers near the experimental curves.  The thin lines represent the predicted
power-laws characterizing the homogeneous regime (see text).  A sizable deviation which scale
the sample size corresponds to sample-sized cluster of fractal geometry.  Large triangles were
used to extract the $R_{c}(L)$ scaling in Inset to Fig.3.  Inset:  Temperature dependence of the
very onset of anomalous dissipation in the thinnest ($19.7 \mu m$) sample.}  
\label{f2}
\end{center} 
\end{figure}

By studying a thickness dependence in appropriate samples we prove now the presence of
geometrical constraints of fractal nature in initial dissipation of HTS in a more quantitative
way (the results on composite tapes, Fig.\ \ref{f1}, represent just a qualitative indication).
The presence of many spurious and/or overlapping effects \cite{fus1} in composite tapes and HTS
films precludes obtaining a firm quantitative information on critical behavior from these
samples.  We performed therefore the measurements of I-dV/dI characteristics on a
well-characterized, non-textured (i.e., isotropic) polycrystalline RBa$_{2}$Cu$_{3}$O$_{7-x}$
(R=Y,Gd) bulk sample (a WLN prototype!)  in many successive steps, after its thickness had been
gradually reduced by fine plan-parallel grinding.  In that way, apart from various thickness,
all the measurements were performed on the same initial sample.  The transport properties of the
sample (e.g., room temperature resistivity, T$_{c}$, resistive transition width) did not change
in all stages of its thickness.  The measurements we report on in this paper covered the sample
thickness range of 20-1000 $\mu m$ (factor of 50).  For thicknesses above approximately 60 $\mu
m$ only the unique `thermodynamic' critical current, accompanied by the usual power-law-like ($
\propto (I-I_{c})^{x}, x\approx 2$) growth of dissipation (specific for scaling regime in a very
large sample, $ \xi \ll L $), have been detected.  In the sample stages involving all smaller
thicknesses the two characteristic currents, $I_{on}$ and $I_{c}$, have been observed, just as
in composite tapes.  Some of the experimental $dV/dI(L)$ curves were shown in Fig.\ \ref{f2}
using moderate (main figure) and a very high dynamical resistance resolution (inset).  The
anomalous dissipative range between $I_{on}$ and $I_{c}$ is rather complex but systematically
depends on sample thickness:  The size of the onset anomaly drastically increases by decreasing
$L$.  The analysis and interpretation have been performed inside the DB model \cite{sust,pre1}
which provides both the limiting behavior in the correlation length scaling range $ \xi \ll L$
(thin lines in Figs.\ \ref{f1}, \ref{f2}) and the estimate of the width of the range of sample
size scaling ($ \xi \geq L$).  The crossover between these two ranges takes place when the
diverging $ \xi$, $ \xi(p)=\ell |p-p_{c}|^{-\nu}$, becomes equal or higher than the sample size.
The unit of length is $\ell$, the network unit cell size (for $\ell$ we used $\ell \approx 5 \mu
m$, the average grain size of the sample).  
In other words, as long as the fraction of `good' bonds deviates from
$p_{c}$ (percolation threshold of an infinite system) by $ \Delta p=(\ell/L)^{1/
\nu}=(\ell/L)^{1.136}$ or less, the macroscopic properties should have a weak dependence on $p$
(i.e., on current in our case) and the underlying ramified sample-sized cluster should be,
geometrically, a fractal.

In DB model \cite{pre1,sust} a linear $p(I)$ approximation has been shown to work well
close to $p_{c}$ (but still outside $\Delta p $):
$p_{c}-p=(c_{2}/c_{1})(p_{c}/I_{c})(I-I_{c})$ where $c_{2}/c_{1}$ ($ \equiv g $
henceforth) is a geometrical factor of order 1.  The current interval compatible with
$\Delta p$, $\Delta I$, reads therefore $\Delta I=I_{c}g( \Delta p
/p_{c})=(I_{c}g/p_{c})(\ell /L)^{1.136}$.  The onset current can be now defined as
$I_{on}=I_{c}- \Delta I$ and the corresponding current density is expected to depend
strongly on sample thickness $L$ :  $J_{on}=J_{c}(1-(g/p_{c})(\ell /L)^{1.136}$).
Experimentally, while the determination of $I_{on}$ in thin samples is quite
straightforward (the onset of dynamical resistance is very sharp, inset to Fig.\ \ref{f2}) the
determination of $I_{c}$ is not; $I_{c}$ represent just a parameter in DB model for
dissipation \cite{pre1}, $dV/dI=R_{f}(gp_{c}/I_{c})^{x}(I-I_{c})^{x}$, where $R_{f}$
represents the total resistance of WLN in homogeneous regime.

\begin{figure}[h!!!]
\begin{center}
\leavevmode
\epsfig{file = 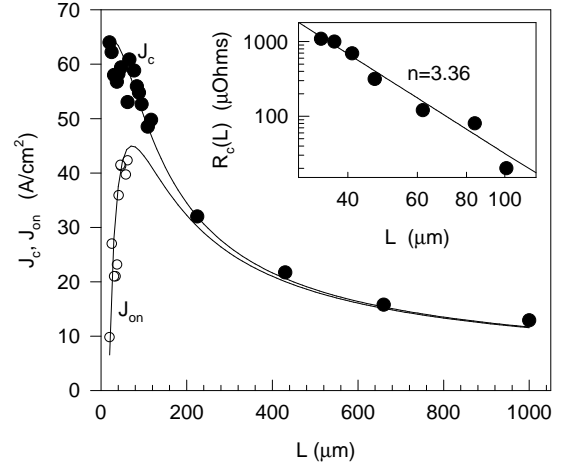, width =0.9 \linewidth, clip=}
\caption[delfsa]{Thickness dependence of thermodynamical critical current ($J_{c}$) and onset
current ($J_{on}$) density- experimental points (symbols) and model predications (thin
lines).  Inset:  Thickness dependence of $dV/dI$ at $I_{c}$ (at criticality) indicates
the fractal geometry of the network.  The slope for homogeneous (Euclidean) electrical
networks is $n=1$.}
\label{f3}
\end{center}
\end{figure}

An interesting observation is that the thermodynamic critical current density $J_{c}$ is
strongly thickness dependent as well, Fig.\ \ref{f3}.  The presence of a size-effect in $J_{c}$
is, however, a rather well-known \cite{der} although somewhat neglected phenomenon.  We found
that the observed increase of $J_{c}$ for a factor of 5 by thinning the cross-section fits
perfectly the general breakdown formula \cite{dux}, $J_{c} \propto 1/(1+(alog_{10}L)^{\alpha})$
which predicts that the critical current density vanishes in thermodynamic limit.  Our results
are compatible with $J_{c}(L)=5J_{c}(L_{max})/(1+(log_{10}(L/L_{min}))^{3})$ where $L_{max}$ and
$L_{min}$ are maximal and minimal sample thickness, respectively.  Inserting the latter
expression into the derived one for $J_{on}$ (comprising only one adjusting parameter, $g$) one
gets the peaked curve, Fig.\ \ref{f3}, as a prediction of this model.  The two dependencies (for
$J_{c}$ and $J_{on}$) joins smoothly in increasing $L$ as $ \Delta p$ (i.e., $ \Delta I$)
continuously vanishes, as well as $ \xi$ itself, in this limit.  The remarkable overlap of
experimental points and the model predictions, Figs.2,3, illustrate the reality of the model, in
spite of its simplicity.  It is interesting to note that Fig.\ \ref{f2} can be interpreted as a
kind of phase diagram:  the thick grey line separates the homogeneous from the fractal phase of
the cluster inside which the initial dissipation grows.

The main quantitative results of this work is plotted as an Inset to Fig.\ \ref{f3}.  It plots
the value of dynamical resistance in fractal phase $R_{c}(I_{c})$ for each available
sample thickness $L$.  The particular current at which $R_{c}$ has been taken was
$I_{c}$, the representative of $p_{c}$ in current-induced transitions.  The $R_{c}(L)$
relationship can be fit nicely by a power-law $R_{c}(L) \propto L^{-n}$.  The exponent
value ($n=3.36$) deviates strongly from the value $n=1$ characterizing homogeneous
networks (the scaling with $n=1$ is strictly obeyed in, e.g., $R_{f}(L)$ dependence).
Also, the quantity $ \nu (n-1)$ which in finite-size scaling calculations gives the
dynamical exponent of homogeneous regime is numerically very close ($=2.1$) to
experimentally well-documented \cite{pre1} value $x=2$ valid in that regime.  Both
arguments provide therefore the evidence for fractal geometry involved in initial dissipation.

The important issue which remains to be clarified is the interpretation of the value of the
exponent $n$, as well as of the related exponent $x$ characterizing the homogeneous regime of
I-V characteristics.  The experimental values of exponents ($x \approx 2, n \approx 3.4$) are,
while mutually consistent, in clear disagreement with those obtained by identifying RSN and
WLN, i.e., $x \approx s \approx 0.8$, $n=s/ \nu +1 \approx 1.9$.  There could be several
reasons why the WLN exponent, $x$, may differ from the classical one, $s$.  A well-known
example is the `Swiss-cheese' morphology in continuum percolation \cite{hal}, equivalent to
the case of broad distribution in bond resistances, shown to influence the exponent.  The
other is experimentally documented non-universal conduction in carbon-black-polymer composite
\cite{bal} attributed to peculiarities of tunneling as a mechanism of local conduction.  Both
the broad distribution and tunneling seems as natural possible causes for exponent deviation
in WLN of HTS.  We also note that there are some obvious differences between current-generated
(WLN) and random-generated (RSN) clusters.  Better understanding of these differences could
probably come from very recent and exciting studies of self-organized \cite{chr} and
`small-world' \cite{wat} networks.

The observation of a fractal dissipative regime offers an interesting new guideline
towards understanding the intrinsic pairing interaction in HTS.  There are namely numerous
arguments that the intrinsic intra- and inter-plane charge transport takes place
actually in a heterogeneous conductive medium \cite{mih}, with percolation playing
probably the important role as well \cite{mes}.  Moreover, these conditions are
considered, according to some authors \cite{phi}, as substantial ingredients of the
mechanism of superconductivity itself.  The involved heterogeneity may rely either on
charge separation, stripes, wires, etc.,\cite{mih}, cluster formation \cite{mes} or on
filamentary fragmentation \cite{phi}.  Under these circumstances it seems quite
reasonable to assume that the intrinsic current transfer may include the fractal
network as well (at least in certain range of relevant transport parameters).  Given
the electrically heterogeneous local properties, combined with vicinity of
metal-insulator transition, the associated elastical (vibrational) network (which is
formally isomorphic to its electrical counterpart
\cite{bun}) might be not only heterogeneous but may posses a fractal geometry as well.
In a fractal elastical lattice the vibrations are, instead of extended phonons, the
localized high-frequency fractons \cite{orb,fus2} which may contribute in pairing.
Relaying on peculiarities of the fracton density of state \cite{orb}, such as high
cut-off frequency and/or high-frequency `missing modes' \cite{orb}, the pairing
temperatures could be higher than those associated to classical phonons.  On basis of
our results we suggest therefore consideration of a fractal dynamical lattice as a
possible source of non-standard pairing interactions at high temperatures.

The author acknowledges fruitful exchange of ideas with J.C.Phillips and is also
indebted to K.Uzelac and I. \v{Z}ivkovi\'{c} for numerous discussions and to
D.Pavuna for his continuous interest and support.  I am also grateful to
P.Kov\'{a}\v{c} and F.C.Matacotta for providing me with samples.


\begin{references}
\bibitem{fus}The closeness of $x$ to the value of conductivity exponent ($t \approx 2$)
inspired some speculations (Ref.\onlinecite{pre1}) about actual identity of these
exponents.  We are however unaware of a convincing general argument equating $x$ with
$t$ at present.
\end{references}
\end{document}